\documentclass[a4paper]{jpconf}
\usepackage{graphicx}
\begin{document}
\title{Search for $^{28}$Si cluster states through the $^{12}$C+$^{16}$O radiative capture }

\author{S. Courtin, A. Goasduff, F. Haas, D. Lebhertz}
\address{Institut Pluridisciplinaire Hubert Curien, UMR 7178, Universit\'e de Strasbourg / CNRS-IN2P3. 23, rue du Loess, 67037 Strasbourg, France}
\ead{sandrine.courtin@iphc.cnrs.fr}

\author{D.G. Jenkins}
\address{Department of Physics, University of York, Heslington, York YO10 5DD, United Kingdom}

\author{D.A. Hutcheon, C. Davis, C. Ruiz}
\address{TRIUMF, Vancouver, Canada V6T 2A3}

\begin{abstract}
The $^{12}$C+$^{16}$O resonant radiative capture reaction has been studied at 5 bombarding energies between E$_{lab}$ = 15.4 and 21.4 MeV, around the Coulomb barrier, at the Triumf laboratory (Vancouver, Canada) using the Dragon 0$^\circ$ spectrometer and the associated BGO array. The most remarquable result is the previously unobserved decay path through $^{28}$Si doorway states of energies around 12 MeV leading to the measurement of new capture cross-sections. The feeding of specific, deformed states in $^{28}$Si from the resonances is discussed, as well as the selective feeding of 1$^+$ T=1 states around 11 MeV.
\end{abstract}

\section{Introduction}
The occurence of $\alpha$ clusters in light nuclei is a well established phenomenon known to occur close to their $\alpha$-decay thresholds, as described by Horiuchi and Ikeda in the late 1960s \cite{ike}. As examples of light cluster nuclei, we shall mention here briefly the $^8$Be nucleus, which ground state (g.s.) decays into 2 $\alpha$ particles and the $^{12}$C sub-closed shell nucleus which first excited 0$^+$ state is located at 7.65 MeV, near the $^8$Be+$\alpha$ fusion threshold. This state, that has been identified by Hoyle to play a major role in the nucleosynthesis of  $^{12}$C, is not described by shell-model calculations but rather by cluster models. The identification of the cluster band built on this 'Hoyle' state is still a challenging goal for experimentalists. In $^{16}$O a clear $alpha$ cluster band has been identified based on a 0$^+$ excited state at 6.05 MeV, a 2$^+$ state at 6.92 MeV and a 2$^+$ state at 10.36 MeV, with large reduced transition probabilities B(E2)(2$^+$ to 0$^+$)= 27 W.u. and B(E2)(4$^+$ to 2$^+$)= 65 W.u. A shell-model description of this very deformed band requires 4p-4h (4 particle-4 hole) excitations to be taken into account.\\
What heavier ions are concerned, in the $sd$ shell, in addition to $\alpha$ clusters,	 $^{12}$C and $^{16}$O are the elementary bricks for the occurence of cluster structures in $^{24}$Mg($^{12}$C-$^{12}$C) and $^{28}$Si($^{12}$C-$^{16}$O). One of the most striking results obtained in the early studies of heavy-ion collisions is the observation of resonant structures in the reaction cross-sections (elastic, inelastic and fusion channels) of some light heavy-ion systems. These structures were found especially in the $^{12}$C+$^{12}$C, $^{12}$C+$^{16}$O and $^{16}$O+$^{16}$O systems \cite{alm,sto,erb}. The resonances are the most pronounced for $^{12}$C+$^{12}$C and, at the Coulomb barrier, they have been found to be narrow ($\Gamma$ $\sim$ 200 keV) and correlated in several reaction channels for $^{12}$C+$^{12}$C and $^{12}$C+$^{16}$O. Due to their small widths, these structures have been interpreted in terms of cluster configurations in $^{24}$Mg($^{12}$C-$^{12}$C) and $^{28}$Si($^{12}$C-$^{16}$O) and have been the subject of numerous experimental studies. Nevertheless, $\gamma$ transitions between these presumably molecular states have never been identified.
We have started recently an experimental programme to search for $\gamma$ signatures of these very deformed states in $^{24}$Mg and $^{28}$Si through the study of the $\gamma$-decay of the resonant radiative capture reactions $^{12}$C($^{12}$C,$\gamma$)$^{24}$Mg and $^{12}$C($^{16}$O,$\gamma$)$^{28}$Si around the Coulomb barrier, where cluster bands with $^{12}$C-$^{12}$C and  $^{12}$C-$^{16}$O structure are predicted to start in $^{24}$Mg and $^{28}$Si respectively \cite{ichi,bay}. 
Radiative capture in these systems had already been investigated at the beginning of the 1980s by the Brookhaven group of Sandorfi {\it et al.} \cite{sand,nath}. In these experiments, excitation functions have been measured using a single large NaI $\gamma$-detector. Due to the piling-up of low energy $\gamma$-rays coming from the much more intense fusion-evaporation channels, the detection was limited to large energy $\gamma$-rays feeding low-lying states in $^{24}$Mg and $^{28}$Si with excitation energies less than $\sim$4.5 MeV. The measured capture cross sections to the first members of the g.s. bands show a series of narrow resonances in both cases. After this radiative capture campaign, Sandorfi {\it et al.}. improved their experimental set-up by using a Wien filter at $\sim$3 $^{\circ}$ in a $^{12}$C+$^{16}$O experiment to select the $^{28}$Si compound nucleus. They could thus measure the feeding of states up to $\sim$9.5 MeV and found, in particular a first indication of the feeding of intermediate states at $\sim$7 MeV in $^{28}$Si from  the resonance \cite{col}.\\
In our experimental campaigns, we have made use of the Dragon $\sim$0 $^{\circ}$ recoil spectrometer and the associated BGO array to measure for the first time the full $\gamma$-decay spectrum in the 
$^{12}$C+$^{12}$C and $^{12}$C+$^{16}$O reactions and to determine the spin and parities of the resonances from the electromagnetic transitions characteristics.
For the $^{12}$C($^{12}$C,$\gamma$) experiment performed at Triumf (Vancouver, Canada), five energies have been used: four on resonance at E$_{c.m.}$=6.0, 6.8, 7.5 and 8 MeV and one off resonance at 6.4 MeV. For the first time in the study of the capture process, the relative strength of all the decay branches could be extracted. We found an important new decay branch to $^{24}$Mg states just above the $\alpha$-threshold (∼ 9.3 MeV) which represents $\sim$50\% of the total $\gamma$ decay width. We were able to assign spin 0$^+$ or 2$^+$ for the resonances at 6.0 and 6.8 MeV, spin 4$^+$ for the resonances at 7.5 and 8 MeV \cite{jenk}. The $^{12}$C($^{16}$O,$\gamma$)$^{28}$Si experiments and results will be discussed below.

\section{Experimental details of the $^{12}$C($^{16}$O,$\gamma$)$^{28}$Si campaign at Triumf}
\begin{center}
\begin{table}[htbc]
\caption{\label{table1} Energies (in MeV) explored in the  $^{12}$C+$^{16}$O experiments at Triumf.}
\centering
\begin{tabular}{@{}*{7}{lll}}
\br
E$_{c.m.}$&E$_{lab}$&E$^*$($^{28}$Si)\\
\mr
6.6 & 15.4 & 23.4 \\
7.2 & 16.8 & 24.0 \\
8.5 & 20.0 & 25.3 \\
8.8 & 20.6 & 25.6 \\
9.0 & 21.1 & 25.8 \\
\br
\end{tabular}
\end{table}
\end{center}

Two experiments have been performed at energies around the Coulomb barrier (E$_{CB}$$\sim$7.8 MeV). The first one used $^{16}$O beam energies at E$_{lab}$ = 19.8, 20.5 and 21.0 MeV corresponding to resonances in the $^{12}$C+$^{16}$O excitation function as measured by Sandorfi {\it et al.} \cite{sand,col} at E$_{c.m.}$ = 8.5 and 9 MeV and to a minimum in this function at 8.8 MeV. The second one was performed at 2 energies, below the Coulomb barrier, E$_{c.m.}$ = 7.2 MeV also corresponding to a maximum in the Sandorfi excitation function and E$_{c.m.}$ = 6.6 MeV, below all previously measured radiative capture data, corresponding to the maximum of a broad structure in the fusion excitation function \cite{chris}. This experiment was performed at subbarrier energies to reduce the number of open fusion-evaporation channels and the range of angular momenta involved, and thus, favor the observation of the radiative capture channel. Energies explored in the full campaign are summerized in Table \ref{table1}. Recoiling nuclei were detected in a double-sided silicon strip detector (DSSSD) at the focal plane of the Dragon spectrometer \cite{hutch} and the radiative capture channel was selected by setting a gate on the time-of-flight between the DSSSD and the $\gamma$ rays in a 30 BGO detector-array. The use of the Dragon spectrometer allowed a rejection factor of the incident beam above 10$^{12}$. This radiative capture experiment is very challenging since the process is dominated by fusion-evaporation which cross-section is 5 orders of magnitude larger than the radiative capture cross-section. Details about the experimental setup and the sort of the data can be found in \cite{lebh} which concerns the energy points above the Coulomb barrier. \\
To obtain quantitative results, like radiative capture cross-sections and to propose spin assignments for the entrance-channel resonances, we have performed Monte Carlo GEANT simulations of the full response function of the experimental setup for our $^{12}$C+$^{16}$O experiments.

\section{Results and discussion}
\begin{figure}
\begin{center}
\includegraphics[width=11.5cm]{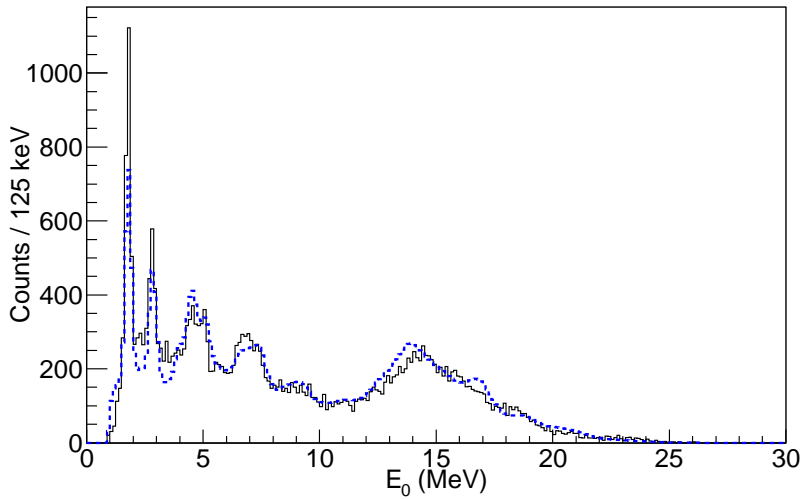}
\end{center}
\caption{\label{fig1} (Color online) Reaction $^{12}$C($^{16}$O,$\gamma$)$^{28}$Si, E$_{c.m.}$ = 9 MeV: highest energy $\gamma$-ray sepctrum compared with numerical simulations for a 6$^+$ entrance resonance (dashed blue).}
\end{figure}

\begin{figure}
\begin{center}
\includegraphics[width=11.5cm]{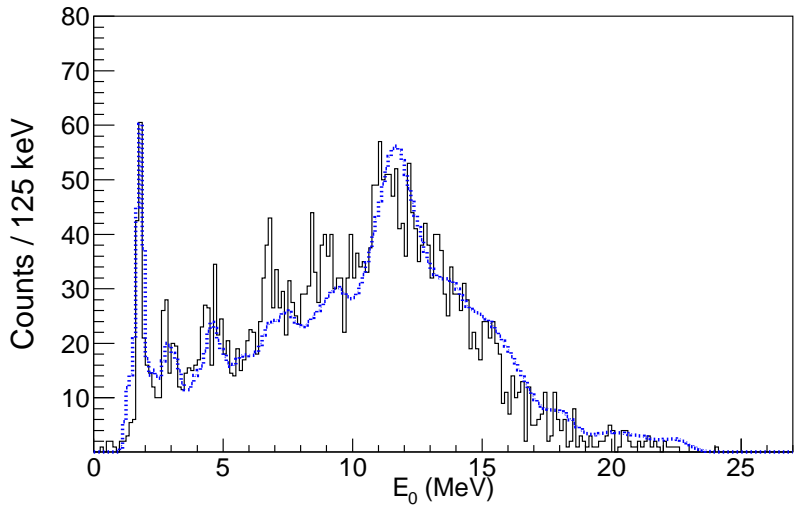}
\end{center}
\caption{\label{fig2} Reaction $^{12}$C($^{16}$O,$\gamma$)$^{28}$Si, E$_{c.m.}$ = 6.6 MeV: highest energy $\gamma$-ray spectrum compared with numerical simulations for a 2$^+$ entrance resonance and enhanced feeding of J$^{\pi}$ = 1$^+$, T = 1 states (dashed blue).}
\end{figure}

Figure \ref{fig1} shows the experimental $\gamma$ spectrum of the highest energy $\gamma$-ray emitted in the cascade (E$_0$) along with numerical simulations (see below) taken at E$_{c.m.}$ = 9 MeV, the highest energy explored in this study. Figure \ref{fig2} shows the same spectrum at  E$_{c.m.}$ = 6.6 MeV, the lowest explored subbarrier energy. The most striking feature of the $\gamma$ spectra measured in these radiative capture experiments is the previously unobserved large bump, around E$_0$ = 12 and 14 MeV respectively, showing the feeding of intermediate doorway states in the decay of the entrance resonance. Using these $\gamma$ data and simulations of the transmission through Dragon, we have measured the radiative capture cross-sections ($\sigma _{RC}$) at the five explored energies. Results for  E$_{c.m.}$ = 9 and 6.6 MeV are reported in Table \ref{table2} along with fusion cross-sections ($\sigma _{fus}$) and previous results from Sandorfi et al. \cite{sand,col}. Capture cross-sections measured in this work are usually much larger than what was previously known due to new decay paths via intermediate $^{28}$Si states. The same feature was already observed in our previous $^{12}$C+$^{12}$C radiative capture studies \cite{jenk}.\\

\begin{center}
\begin{table}[htbc]
\caption{\label{table2} Cross-sections.}
\centering
\begin{tabular}{@{}*{7}{llll}}
\br
E$_{c.m.}$ (MeV)&$\sigma _{fus}$ (mb) from \cite{chris,cuj} &$\sigma _{RC}$ ($\mu$b) from \cite{col} & $\sigma _{RC}$ ($\mu$b) this work\\
\mr
6.6 & 8 & - & 0.2(0.03) \\
9.0 & 244 & 3.1 & 23.4(5.7) \\
\br
\end{tabular}
\end{table}
\end{center}

The complete Dragon-BGO array setup geometry has been included in the numerical simulations as well as properties of the $^{28}$Si compound nucleus recoil -i.e. mass number, excitation energy, bound and quasi-bound states from the literature and their branching ratios- to calculate the $\gamma$ spectra. 
The entrance angular momentum is a parameter of these simulations which then take into account the electromagnetic transition properties in the $^{28}$Si mass region \cite{endt}. We have first used coupled-channel calculations to predict a statistical entrance-channel spin distribution. We have noticed that, to reproduce the experimental data, at some energies, a dominant unique spin in the entrance-channel is needed, which is coherent with the population of a resonance in $^{28}$Si. At E$_{c.m.}$ = 9 MeV, we have assigned a spin 6$^+$ to the resonance. The corresponding simulated spectrum is reported in Fig. \ref{fig1} (blue dashed line). At E$_{c.m.}$ = 6.6 MeV, the entrance spin that best fits the data is 2$^+$. These assignments are in very good agreement with a previous study of the breakup of $^{28}$Si into $^{12}$C+$^{16}$O in which spin systematics was established for the breakup band, with a very large molecule-like moment of inertia, down to the Coulomb barrier \cite{met}. \\
Several specific features of the decay have been identified at these two energies :

\begin{itemize}
\item at E$_{c.m.}$ = 9 MeV, where we have attributed a spin 6$^+$ to the resonance, an enhanced feeding of the 4$^+$ member of the prolate band is been observed as well as the direct feeding of the 3$^-$ state at 6.88 MeV, from the resonance. This state is the band head of the K$^{\pi}$ = 3$^-$ octupole band which mainly decays to the $^{28}$Si ground state with a strong E3 transition of reduced width B(E3) = 20 W.u.  This and the fact that the resonance spin is in good agreement with a very deformed breakup band is a good indication of a deformed, presumably molecular, entrance-channel.
\item at E$_{c.m.}$ = 6.6 MeV, where our data suggest a resonance with J$^{\pi}$ = 2$^+$, the phase space open for the decay is smaller, and  we have clearly identified the enhanced decay of the resonance via  J$^{\pi}$ = 1$^+$, T=1 states in $^{28}$Si and in particular the 1$^+$ state at 11.45 MeV which has a large M1 decay width to the ground state and is selectively populated in proton and electron inelastic scattering \cite{lut,cra}. Angular distributions of the $\gamma$-rays in the energy region E$_{0}$ = 11-13 MeV suggest dipole transitions from the resonance at E$^*$($^{28}$Si) = 23.4 MeV to the intermediate states (10-12 MeV) and from the intermediate states to the $^{28}$Si ground state. 
\end{itemize}
At the lowest bombarding energy, it would be of interest to measure the $\gamma$ decay spectrum with better resolution, in particular in the energy region E$_{0}$ = 14-18 MeV (see Fig. \ref{fig2}) to clearly indentify eventual structural effects, as observed above the Coulomb barrier at E$_{c.m.}$ = 9 MeV such as the enhanced feeding of the $^{28}$Si octupole band. This could be achieved with the use of new generation $\gamma$ detector arrays based on scintillators such as LaBr$_3$. Such a project, called PARIS \cite{maj} has started some years ago and we plan to study radiative capture in the $^{12}$C+$^{12}$C and $^{12}$C+$^{16}$O systems using this array in the coming years.

Our studies of $^{12}$C+$^{12}$C and $^{12}$C+$^{16}$O radiative capture reactions have revealed interesting similarities:

\begin{itemize}
\item At the lowest explored energies, E$_{c.m.}$ = 6 MeV for C+C \cite{jenk} and E$_{c.m.}$ = 6.6 MeV for C+O, we have assigned a spin 2$^+$ for the entrance resonance and measured a favoured decay via 1$^+$, T=1 states. 
\item At higher bombarding energies, we have observed the enhanced feeding of specific $^{24}$Mg and $^{28}$Si states, i.e. the prolate ground state band of $^{24}$Mg and the prolate excited band of $^{28}$Si, starting at 6.7 MeV, the ground state of $^{28}$Si being oblate.
\end{itemize}
Previous studies of the $^{12}$C($^{16}$O,$\gamma$)$^{28}$Si reaction at higher energies, aiming at measuring the isospin mixing in the $^{28}$Si excitation energy region around 35 MeV, discussed the reaction mechanism in terms of populating the giant dipole resonance (GDR) built on excited states \cite{har}. It might be that in our studies, the giant quadrupole resonance (GQR) is excited, built on the $^{24}$Mg prolate ground state for $^{12}$C+$^{12}$C \cite{cujecf} and on the prolate excited band in the case of $^{12}$C+$^{16}$O. In fact, in heavy-ion radiative capture reactions Coulomb barriers are high, the entrance energy in the coumpound nucleus is thus rather large, $\sim$ 25 MeV in the present study. This excitation energy falls right in the range of giant resonances. More precisely, the experimental maximum of the GQR in $^{28}$Si is around 18.8 MeV \cite{young}, the GQR built on the prolate band of $^{28}$Si (starting at 6.7 MeV) would thus lie around 25.5 MeV.

\end{document}